# New SpiroPlanck Heuristics for High Energy Physics Networking and Future Internet Testbeds

M. A. El-Dosuky[1], M. Z. Rashad[1], T. T. Hamza[1], and A.H. EL-Bassiouny[2]

[1] Department of Computer Sciences, Faculty of Computers and Information sciences, Mansoura University, Egypt

**mouh_sal_010@mans.edu.eg**

**magdi_12003@yahoo.com**

**Taher_Hamza@yahoo.com**

[2] Department of Mathematics, Faculty of Sciences, Mansoura University, Egypt

**el_bassiouny@mans.edu.eg**

## ABSTRACT
The need for data intensive Grids, and advanced networks with high performance that support our science has made the High Energy Physics community a leading and a key co-developer of leading edge wide area networks. This paper gives an overview of the status for the world's research networks and major international links used by the high energy physics and other scientific communities, showing some Future Internet testbed architectures, scalability, geographic scope, and extension between networks. The resemblance between wireless sensor network and future internet network, especially in scale consideration as density and network coverage, inspires us to adopt the models of the former to the later. Then we test this assumption to see that this provides a concise working model. This paper collects some heuristics that we call them SpiroPlanck and employs them to model the coverage of dense networks. In this paper, we propose a framework for the operation of FI testbeds containing a test scenario, new representation and visualization techniques, and possible performance measures.  Investigations show that it is very promising and could be seen as a good optimization

## 1. INTRODUCTION
The need for global collaborations, data intensive Grids, and advanced networks with high performance that support our science has made the High Energy  Physics (HEP) community a leading and a key co-developer of leading edge wide area networks. The major cooperation efforts, such as CMS[1] and ATLAS [2] who are building experiments for CERN's [3] Large Hadron Collider (LHC) [4]. HEP requires building data Grids capable of sharing and processing massive physics datasets, rising from the Petabyte ($10^{15}$ byte) to the Exabyte ($10^{18}$ byte) scale. A roadmap for HEP networks in the coming decade [5] illustrates gradual raises every 2-3 years.

With more and more emphasis that the current Internet architecture is being challenged by unprecedented overwhelming scale, while we increasingly rely on it for all aspects of our lives. The ability to evolve the Internet architecture requires new proposed protocols with the ability to test them. Testing these protocols requires testbeds that have significant scale, geographic scope, and that have a programmable Internet waist to permit experimentation with new addressing, forwarding, routing, and signaling paradigms.

This paper surveys the global research networks, the major international links used by the HEP, and the state-of-art of the advancement happening in Future Internet (FI) testbeds.  We propose a framework for the operation of FI testbeds containing a test scenario, new representation and visualization techniques, and possible performance measures.

Section 2 reviews new technologies and the major network backbones and international links used by HEP, showing some FI testbed architectures, scalability, geographic scope, and extension between networks such as NLR, GÉANT2, JANET, and KOREN. Section 3 introduces the proposed framework. Section 4 evaluates the proposed framework. Section 5 concludes the paper.

## 2. STATE-OF-ART
Dense wavelength division multiplexing (DWDM) uses several light wavelengths to transmit signals over a single optical fiber allowing new services to be provisioned over existing infrastructure [6]. Shifting to DWDM by adding dynamically provisioned optical paths (also called "lambdas") to build "Lambda Grids" ([7], [8]). The Global Lambda Integrated Facility (GLIF) [9] is promotes the shift to lambda networking by supporting data-intensive network traffic and middleware development.

The major backbones with 10 Gbps wavelengths are National Lambda Rail (NLR) [10] in USA, CA*net 4 [11] in Canada, and GEANT [12] is pan-European connecting several major networks in European countries, remarkably SuperJANET4 [13], RENATER [14], G-WIN [15], GARR-G [16], SURFNet [17], PIONIER [18], SANET [19], CESnet [20], and NORDUnet [21]. Korea has KREONET [22] and KOREN [23]. Japan has SINET/SuperSINET [24] and JGN2 testbeds. China has CSTNET [25] and CERNET [26]. Outside these regions, network connections are slower and in an increasing danger of being left behind [27]. One hopeful attempt is EUMEDCONNECT [28] to overcome the humble Euro-Mediterranean connectivity.

## 3. PROPOSED FRAMEWORK

### 3.1 Modeling Network Coverage

The resemblance between wireless sensor network[29] and future internet network, especially in scale consideration as density and coverage, inspires us to adopt the models of the former to the later. Then we will test this assumption to see if this provides a concise working model.

Density $\lambda$ is a parameter to fulfill network coverage with energy saving [30]. It can be calculated as the average number of neighbors per node:

$$\lambda = \frac{N\pi R^2}{A} \quad (1)$$

where $R$ is node range, $A$ is area of sensor field, and $N$ is total number of deployed nodes.

The network has a very low probability of isolated node [31]:

$$P(isolated\ node) = (1 - e^{-\lambda})^N \quad (2)$$

Connectivity is analyzed to show if the probability of isolation is enough to represent connectivity [32].

### 3.2 Visualizing Multiple Lambdas

Visualizing multiple lambdas is a new concept and may require a new periodic representation. We recommend Spirograph [33], a curve shaped by gently sloping a smaller circle of radius r inside a larger circle or radius R. If the pen offset of the pen point in the moving circle is a, then the equation of the resulting curve at time t is

$$x = (r1+r2)*\cos(t) - (r2+a)*\cos(((r1+r2)/r2)*t)$$
$$y = (r1+r2)*\sin(t) - (r2+a)*\sin(((r1+r2)/r2)*t) \quad (3)$$

A lambda is depicted as a superposed circle with a specific radius. Figure 1 shows a Spirograph with r1=180, r2=40, a=15.

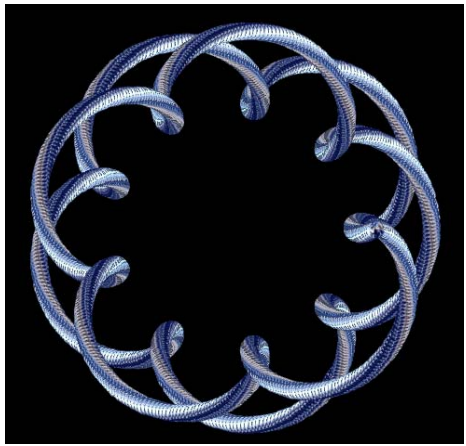

**Fig. 1 sample Spirograph**

### 3.3 New Performance Measures

Seeking a new performance measure that depends on lambdas, we recommend Planck's black-body equations that defines the emitted power as a function in wavelength [34]:

$$P_\lambda = \frac{2\pi hc^2}{\lambda^5(e^{hc/\lambda kT}-1)} \quad (4)$$

where $P_\lambda$ is power per m² area per m wavelength, h is Planck's constant (6.626x10-34 J.s), c is the speed of Light (3x108 m/s), $\lambda$ is the wavelength (m), k is Boltzmann Constant (1.38 x10-23 J/K), T is temperature (K), which may serve the role of SNR. The curve is shown in Figure 2.

The MATLAB code to generate this curve is given below:

```
h = 6.6261*10^-34;
c = 2.9979*10^8;
k = 1.3807*10^-23;
lambda = 1e-9:10e-9:3000e-9;
A1=(h.*c)./(k.* 4500.*lambda);
A2=(h.*c)./(k.* 6000.*lambda);
A3=(h.*c)./(k.* 7500.*lambda);

qu =(8.*pi.*h.*c)./lambda.^5;
c1= qu.*(1./(exp(A1)-1));
c2= qu.*(1./(exp(A2)-1));
c3= qu.*(1./(exp(A3)-1));

plot(lambda, c1,'-.b',lambda, c2, 'g--',lambda, c3, '-ro')

xlabel('Wavelength (\lambda)','FontSize',16)

ylabel('Intensity','FontSize',16)

title('\it{ Wavelength and Intensity}','FontSize',16)

hleg1 = legend('T_1 = 4500','T_2 = 6000', 'T_3 = 7500');
```

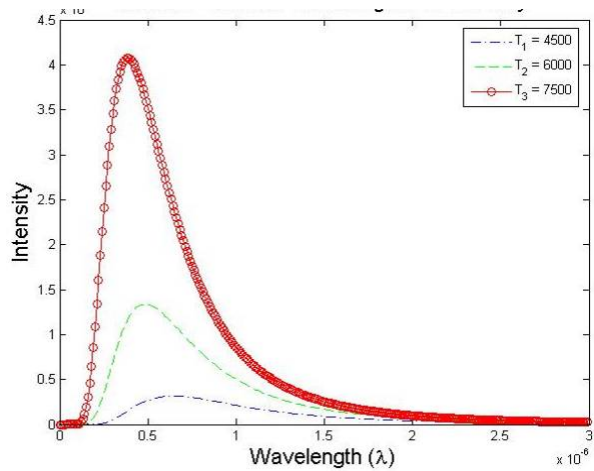

**Figure 2 Relation between Wavelength and Intensity**

## 3.4 SpiroPlanck

Based on these three heuristics, we propose the following algorithm.

---

**function** SpiroPlanck (*Spirograph,* R, A)

**input:**

  *Spirograph*, a set of covering point $(x_i, y_i)$

  R, node range

  A, area of the sensor field,

**Variables:**

  $\lambda$ , density

  $p$ , p(isolated node)

  N, no of deployed nodes, initially =1

  *closed, set of occupied points*, initially empty

  *current, current intersection point*

  *threshold, fixed to 0.1*

  **begin**

$$\lambda = \frac{N\pi R^2}{A} \;,$$

---

```
                    p = (1 - e^{-λ})^N
            while p < threshold do
                current = SELECT(Spirograph)
                    if current ∉ closed then
                        closed = closed ∪ { current }
                        N = N+1
                                 NπR²
                            λ = ─────  ,
                                  A
    p = (1 - e^{-λ})^N
                    end if;
                             2πhc²
                     P_λ = ──────────────
                           λ⁵(e^{hc/λkT} - 1)
                    end while;
        end
```

**Fig. 3 Pseudo code for SpiroPlanck.**

Note that *Spirograph* describes the covering boundary of the network, fed as a set of pairs representing points, or calculated using parameters in Eq. 3.

## 4. EVALUATION
For benchmark analysis, we adopt these parameters [35]: 100x100m area and 7m sensing range. Effective density is:

$$\lambda = \frac{N\pi R^2}{A} \quad (5)$$

The probability that a unit area is in the range of *n* nodes is:

$$P(n) = P_R^n (1 - P_R)^{N-1-n} \binom{N-1}{n} \quad (6)$$

where

$$P_R = \frac{\pi R^2}{L^2} \quad (7)$$

For large values of *N*, this binomial distribution becomes Poisson distribution [36]:

$$P(n) = \frac{e^{-\lambda_s} \lambda^n}{n!} \quad (8)$$

### 4.1 Test Scenario
The following scenario can be easily executed on any free testbed, between any two cities p1 and p2.

1) P1's server generates streaming traffic to #1 switch.
2) Verify the transmission of traffic to P2's Receiver #1.
3) Execute commands in Open flow Controller.
4) Verify the termination of transmission to P2's server #1.
5) Verify the transmission of traffic to P1's Receiver #2.

Setting p1=Seattle in USA, and p2= Daejeon in Korea.

### 4.2 Test Results
One problem to consider is the evaluation criterion for all wired and wireless gateways. Open Shortest Path First (OSPF), a traditional link-state routing protocol for Internet Protocol (IP), can be specified [37], with back-compatibility for current infrastructures [38].

This encourage us to propose the OSPF overhead as one of evaluation measures for comparing algorithms proposed for networks of increasing size. Table 1 and Fig. 4 show the OSPF overhead generated by the OPNET [39].

**Table 1 OSPF overhead**

| Number of nodes | *Pt-to-Mpt* simulation | *Pt-to-Mpt* emulation | SpiroPlanck |
|---|---|---|---|
| 10 | 41 | 19 | 9 |
| 15 | 115 | 63 | 13 |
| 20 | 250 | 109 | 18 |
| 25 | 432 | 204 | 25 |
| 30 | 712 | 458 | 45 |

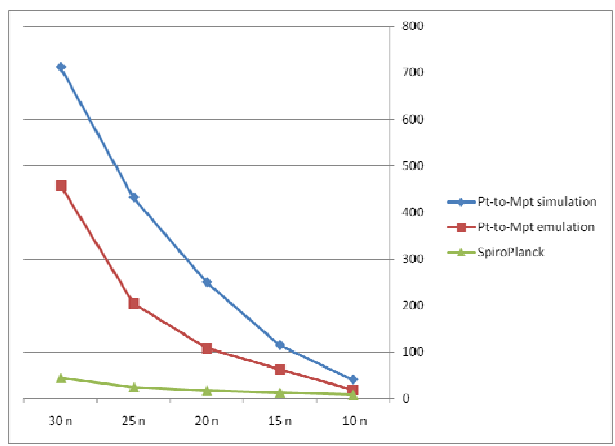

**Fig. 4 OSPF overhead**

Point-to-multipoint *(Pt-to-Mpt)* refers to OSPF's preferred network interface for partial mesh networks. The horizontal axis is for the number of nodes. The vertical axis is for OSPF overhead. Results agree with those stated in [40].

## 5. CONCLUSION

We give an overview of the status for the world's research networks and major international links used by the high energy physics and other scientific communities, showing some Future Internet testbed architectures, and scalability.

The resemblance between wireless sensor network and future internet network, especially in scale consideration as density and network coverage, inspires us to adopt the models of the former to the later. Then we test this assumption to see that this provides a concise working model.

This paper collects some heuristics that we call them SpiroPlanck and employs them to model the coverage of dense networks. In this paper, we propose a framework for the operation of FI testbeds containing a test scenario, new representation and visualization techniques, and possible performance measures. Investigations show that it is very promising and could be seen as a good optimization.

*D950-10897-1*, Available at http://hipserver.mct.phantomworks.org/ietf/ospf/reports/Boeing-D950-10897-1.pdf, 2012.

7